\def\etal{{\it et al.}}
\def\ie{{\it i.e.}}

\def\~{{$\tilde{\phantom{a}}$}}

\documentclass [12pt] {article}
\usepackage{epsfig}
\usepackage{color}

\def\thebibliography#1{\section{References}\markboth
 {REFERENCES}{REFERENCES}\list
 {[\arabic{enumi}]}{\settowidth\labelwidth{[#1]}\leftmargin\labelwidth
 \advance\leftmargin\labelsep
 \usecounter{enumi}}
 \def\newblock{\hskip .11em plus .33em minus -.07em}
 \sloppy
 \sfcode`\.=1000\relax}
\def\upcite#1{\raise6pt\hbox{\scriptsize
\cite{#1}}}
  \def\lsim{\mathrel {\vcenter {\baselineskip 0pt \kern 0pt
    \hbox{$<$} \kern 0pt \hbox{$\sim$} }}}
    \def\gsim{\mathrel {\vcenter {\baselineskip 0pt \kern 0pt
    \hbox{$>$} \kern 0pt \hbox{$\sim$} }}}

\def\hline{\noalign{\hrule \vskip2pt}}

\def\|{\ifmmode\Vert\else \char`\|\fi}
\ifx\oldzeta\undefined                          
  \let\oldzeta=\zeta                            
  \def\zzeta{{\raise 2pt\hbox{$\oldzeta$}}}     
  \let\zeta=\zzeta                              
\fi

\ifx\oldchi\undefined                           
  \let\oldchi=\chi                              
  \def\cchi{{\raise 2pt\hbox{$\oldchi$}}}       
  \let\chi=\cchi                                
\fi

\def\frac#1#2{{#1 \over #2}}

\def\half{\ifinner {\scriptstyle {1 \over 2}}
   \else {1 \over 2} \fi}

\def\simge{\mathrel{%
   \rlap{\raise 0.511ex \hbox{$>$}}{\lower 0.511ex \hbox{$\sim$}}}}
\def\simle{\mathrel{
   \rlap{\raise 0.511ex \hbox{$<$}}{\lower 0.511ex \hbox{$\sim$}}}}

\def\buildchar#1#2#3{{\null\!                   
   \mathop#1\limits^{#2}_{#3}                   
   \!\null}}                                    
\def\overcirc#1{\buildchar{#1}{\circ}{}}

\def\slashchar#1{\setbox0=\hbox{$#1$}           
   \dimen0=\wd0                                 
   \setbox1=\hbox{/} \dimen1=\wd1               
   \ifdim\dimen0>\dimen1                        
      \rlap{\hbox to \dimen0{\hfil/\hfil}}      
      #1                                        
   \else                                        
      \rlap{\hbox to \dimen1{\hfil$#1$\hfil}}   
      /                                         
   \fi}                                         %

\def\subrightarrow#1{
  \setbox0=\hbox{
    $\displaystyle\mathop{}
    \limits_{#1}$}
  \dimen0=\wd0
  \advance \dimen0 by .5em
  \mathrel{
    \mathop{\hbox to \dimen0{\rightarrowfill}}
       \limits_{#1}}}                           

\def\overlay#1#2{\ifmmode%
\setbox0=\hbox{$#1$}%
\setbox1=\hbox to\wd0{\hss$#2$\hss}\else%
\setbox0=\hbox{#1}%
\setbox1=\hbox to\wd0{\hss#2\hss}\fi%
#1\hskip-\wd0\box1 }

\def\pmb#1{\leavevmode\setbox0=\hbox{#1}%
\kern-.02em\copy0\kern-\wd0
\kern.04em\copy0\kern-\wd0
\kern-.02em\raise.04em\box0 }

\def\vereq#1#2{\lower3pt\vbox{\baselineskip1.5pt \lineskip1.5pt
\ialign{$\m@th#1\hfill##\hfil$\crcr#2\crcr\sim\crcr}}}

\def\tensor#1{\protect\@ontopof{#1}{\leftrightarrow}{1.15}\mathord{\box2}}
\def\overstar#1{\protect\@ontopof{#1}{\ast}{1.15}\mathord{\box2}}
\def\overdots#1{\protect\@ontopof{#1}{\cdots}{1.0}\mathord{\box2}}
\def\overcirc#1{\protect\@ontopof{#1}{\circ}{1.2}\mathord{\box2}}
\def\loarrow#1{\protect\@ontopof{#1}{\leftarrow}{1.15}\mathord{\box2}}
\def\roarrow#1{\protect\@ontopof{#1}{\rightarrow}{1.15}\mathord{\box2}}

\def\@ontopof#1#2#3{%
{\mathchoice
{\@@ontopof{#1}{#2}{#3}\displaystyle\scriptstyle}%
{\@@ontopof{#1}{#2}{#3}\textstyle\scriptstyle}%
{\@@ontopof{#1}{#2}{#3}\scriptstyle\scriptscriptstyle}%
{\@@ontopof{#1}{#2}{#3}\scriptscriptstyle\scriptscriptstyle}%
}%
}

\def\@@ontopof#1#2#3#4#5{%
\setbox0=\hbox{$#4#1$}%
\setbox1=\hbox{$#5#2$}%
\setbox2=\hbox{}\ht2=\ht0 \dp2=\dp0 %
\ifdim\wd0>\wd1 %
\setbox1=\hbox to\wd0{\hss\box1\hss}%
\mathord{\rlap{\raise#3\ht0\box1}\box0}%
\else   %
\setbox1=\hbox to.9\wd1{\hss\box1\hss}%
\setbox0=\hbox to\wd1{\hss$#4\relax#1$\hss}%
\mathord{\rlap{\copy0}\raise#3\ht0\box1}%
\fi
}%

\def\lambdabar{\protect\@lambdabar}
\def\@lambdabar{%
\relax
\bgroup
\def\@tempa{\hbox{\raise.73\ht0
\hbox to0pt{\kern.25\wd0\vrule width.5\wd0
height.1pt depth.1pt\hss}\box0}}%
\mathchoice{\setbox0\hbox{$\displaystyle\lambda$}\@tempa}%
{\setbox0\hbox{$\textstyle\lambda$}\@tempa}%
{\setbox0\hbox{$\scriptstyle\lambda$}\@tempa}%
{\setbox0\hbox{$\scriptscriptstyle\lambda$}\@tempa}%
\egroup
}

\def\corresponds{{\lower.2ex\hbox{=}}{\rm\kern-.75em^\triangle}}
\def\succsim{\succ\kern-.9em_\sim\kern.3em}
\def\precsim{\prec\kern-1em_\sim\kern.3em}
\def\slantfrac#1#2{\kern1em^{#1}\kern-.3em/\kern-.1em_{#2}}

\begin{document}
                                                                
\begin{center}
{\Large\bf An Off-Axis Neutrino Beam}
\\

\medskip

Kirk T.~McDonald
\\
{\sl Joseph Henry Laboratories, Princeton University, Princeton, NJ 08544}
\\
(November 6, 2001)
\end{center}

\section{Problem}

A typical high-energy neutrino beam is made from the decay of $\pi$
mesons that have been produced in proton interactions on a target,
as sketched in the figure below.

\vspace{0.1in}
\centerline{\includegraphics[width=5in]{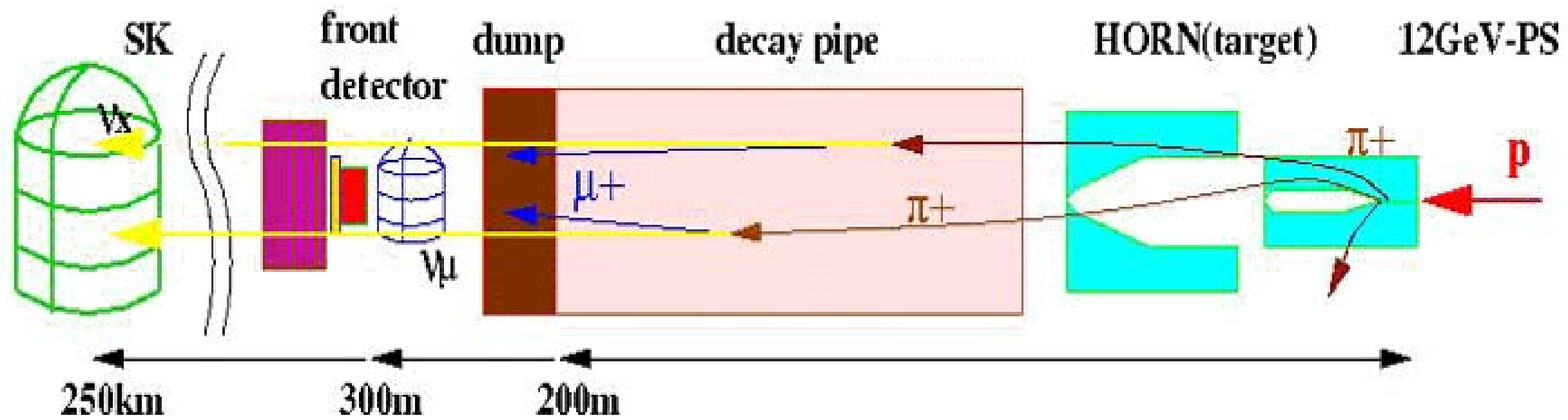}} 

Suppose that only positively charged particles are collected by the
``horn''.  The main source of neutrinos is then the decay $\pi^+ \to
\mu^+ \nu_\mu$.

\begin{enumerate}
\item
Give a simple estimate of the relative number of other types of
neutrinos than $\nu_\mu$ in the beam.

\item
If the decay pions have energy $E_\pi \gg m_\pi$, what is the
 characteristic
angle $\theta_C$ of the decay neutrinos with respect to the direction
of the $\pi^+$?

\item
If a neutrino is produced with energy $E_\nu \gg m_\pi$, what is the maximum
angle $\theta_{\rm max}(E_\nu)$ between it and the direction of its parent pion 
(which can have any energy)?
What is the maximum energy $E_\nu$ at which a neutrino can be produced in the
decay of a pion if it appears at a given angle $\theta$ with respect to
the pion's direction?  

Parts 4 and 6 explore consequences of the existence of these maxima.

\item
Deduce an analytic expression for the energy-angle
spectrum $d^2N / dE_\nu d\Omega$ for neutrinos
produced at angle $\theta \leq \theta_C$ to the proton beam.  You
may suppose that $E_\nu \gg m_\pi$, that the pions are produced
with an energy spectrum
$dN / dE_\pi \propto (E_p - E_\pi)^5,$
where $E_p$ is the energy of the proton beam,
and that the ``horn'' makes all pion momenta parallel to that of the
proton beam.

\item
At what energy $E_{\nu,\rm peak}$ does the neutrino spectrum peak for
$\theta = 0$?

\item
Compare the characteristics of a neutrino beam at $\theta = 0$ with an off-axis
beam at angle $\theta$ such that $E_{\nu,\rm max}(\theta)$ is less than
$E_{\nu,\rm peak}(\theta = 0)$.

\end{enumerate}

Facts: $m_\pi = 139.6$ MeV/$c^2$, $\tau_\pi = 26$ ns, 
$m_\mu = 105.7$ MeV/$c^2$, $\tau_\mu = 2.2\ \mu$s.  In this problem,
neutrinos can be taken as massless.

\newpage

\section{Solution}

In this solution we use units where $c = 1$.

\begin{enumerate}
\item
Besides the $\nu_\mu$ from the decay $\pi^+ \to \mu^+ \nu_\mu$, 
the beam  will also contain $\bar\nu_\nu$ and $\nu_e$ from the
subsequent decay $\mu^+ \to e^+ \nu_e \bar\nu_\mu$.  Both of these
decays occur (primarily) in the ``decay pipe" shown in the figure.
As both the pions and muons of relevance are relativistic in this
problem, they both have about the same amount of time to decay before
they are absorbed in the ``dump''.  Hence, the ratio of number of
muon decays to pion decays is roughly the same as the ration of there
lifetimes, \ie, about 0.01.  Our simple estimate is therefore,
\begin{equation}
{N_{\nu_e} \over N_{\nu_\mu}} = {N_{\bar \nu_\mu} \over N_{\nu_\mu}}
\approx 0.01.
\label{s11}
\end{equation} 

Experts may note that an additional source of $\nu_e$ is the decay
$\pi^+ \to e^+ \nu_e$ at the level of $10^{-4}$.  Also, $K^+$ mesons
will be produced by the primary proton interaction at a rate about
10\% that of $\pi^+$.  About 65\% of $K^+$ decays are to 
$\mu^+ \nu_\mu$, which add to the main $\nu_\mu$ beam, but about 5\%
of the decays are to $\pi^+ \pi^0 \nu_e$, which increases the $\nu_e$
component of the neutrino beam by about $0.1 \times 0.05 = 0.005$.

\item
Parts 2-6 of this problem are based on the kinematics of charged pion
decay, which are closely related to kinematic features of neutral pion
decay, $\pi^0 \to \gamma \gamma$ \cite{Sternheimer}.

Experts may guess that the characteristic angle of the decay neutrinos
with respect to the parent pion is $\theta_C = 1/\gamma_\pi 
= m_\pi / E_\pi$.  The details of the derivation are needed in
part 3.

We consider the decay $\pi \to \mu \nu$ in the rest frame of the pion
(in which quantities will be labeled with the superscript $\star$) and
transform the results to the lab frame.

Energy-momentum conservation can be written as the 4-vector relation,
\begin{equation}
\pi = \mu + \nu,
\label{s1}
\end{equation}
where the squares of the 4-vectors are the particle masses,
$\pi^2 = m_\pi^2$, $\mu^2 = m_\mu^2$ and $\nu^2 = 0$.  As we are
not concerned with details of the muon, it is convenient to rewrite
eq.~(\ref{s1}) as
\begin{equation}
\mu = \pi - \nu,
\label{s2}
\end{equation}
and square this to find
\begin{equation}
m_\mu^2 = m_\pi^2 - 2(\pi \cdot \nu).
\label{s3}
\end{equation}
In the rest frame of the pion, its 4-vector can be written
\begin{equation}
\pi = (m_\pi,0,0,0).
\label{s4}
\end{equation}
Taking the $z$ axis to be the direction of the pion in the lab frame,
the 4-vector of the (massless) neutrino in the pion rest frame can be
written as
\begin{equation}
\nu = (E_\nu^\star,E_\nu^\star \sin\theta^\star,0,
E_\nu^\star \cos\theta^\star),
\label{s5}
\end{equation}
since the energy and momentum of a massless particle are equal.
The 4-vector product $(\pi \cdot \nu) = \pi_0 \nu_0 - \pi_i \nu_i$ 
is therefore
\begin{equation}
(\pi \cdot \nu) = m_\pi E_\nu^\star.
\label{s6}
\end{equation}
Hence, from eq.~(\ref{s3}) the energy of the neutrino in the pion
rest frame is
\begin{equation}
E_\nu^\star = {m_\pi^2 - m_\mu^2 \over 2 m_\pi} = 29.8\ \mbox{MeV},
\label{s7}
\end{equation}
using the stated facts.

We can now transform the neutrino 4-vector (\ref{s5}) to the lab frame,
using the Lorentz boost $\gamma_\pi = E_\pi / m_\pi$,
\begin{eqnarray}
\nu & = & (E_\nu,E_\nu \sin\theta,0,E_\nu \cos\theta)
\nonumber \\
& = & (\gamma_\pi E_\nu^\star(1 + \beta_\pi \cos\theta^\star),
E_\nu^\star \sin\theta^\star,0,
\gamma_\pi E_\nu^\star(\beta_\pi + \cos\theta^\star)).
\label{s8}
\end{eqnarray}

The pion has spin zero, so the decay is isotropic in the pion rest
frame.  A relation  for the angle $\theta$ between the neutrino and its parent pion
can be obtained from the 1 and 3 components of eq.~(\ref{s8}),
\begin{equation}
\tan\theta = { E_\nu^\star \sin\theta^\star \over \gamma_\pi E_\nu^\star
(\beta_\pi + \cos\theta^\star)}\, .
\label{s14b}
\end{equation}
The characteristic angle of the decay in the lab frame is
usefully associated with decays at $\theta^\star = 90^\circ$ in the
pion rest frame.  Thus,
\begin{equation}
\tan\theta_C = {1 \over \gamma_\pi \beta_\pi}.
\label{s9}
\end{equation}
When $E_\pi \gg m_ \pi$ then $\gamma_\pi \gg 1$, $\beta_\pi \approx 1$,
and 
\begin{equation}
\theta_C \approx {1 \over \gamma_\pi} = {m_\pi \over E_\pi} \ll 1.
\label{s10}
\end{equation} 

\item
We now consider the lab angle (\ref{s14b}) between the neutrino and its parent
pion with emphasis on the neutrino energy rather than the pion energy.
If $E_\nu \gg m_\pi$, then $E_\pi \gg m_\pi$ also, so $\gamma_\pi \gg 1$
and $\beta_\pi \approx 1$.  Then we can write
\begin{equation}
\tan\theta
\approx { E_\nu^\star \sin\theta^\star \over \gamma_\pi E_\nu^\star
(1 + \cos\theta^\star)} 
\approx { E_\nu^\star \sin\theta^\star \over E_\nu} \, ,
\label{s14}
\end{equation}
using the time component of eq.~(\ref{s8}).
Since $\sin\theta^\star$ cannot exceed unity, we see that there is a maximum
lab angle $\theta$ relative to the direction of the pion at which a neutrino 
of energy $E_\nu$ can appear, namely
\begin{equation}
\theta_{\rm max} \approx {E_\nu^\star \over E_\nu}
\approx {30\ {\rm MeV} \over E_\nu}\, ,
\label{s14a}
\end{equation}
which is small for $m_\pi \ll E_\nu$.

If instead, the angle $\theta$ is given, eq.~(\ref{s14}) also tells us that
\begin{equation}
E_\nu \approx { E_\nu^\star \sin\theta^\star \over \tan\theta} 
\leq { E_\nu^\star \over \tan\theta}\, .
\label{s14c}
\end{equation}

\item
We desire the neutrino spectrum in terms of the laboratory quantities
$E_\nu$, $\theta$ and $\phi$.  We expect that the spectrum is uniform
in the azimuthal angle $\phi$.  We are given the energy spectrum
$dN / dE_\pi \propto (E_p - E_\pi)^5$ of the parent pions, 
and we have deduced that the spectrum
is isotropic in the pion rest frame, \ie, flat in $\cos\theta^\star$.
Hence, we seek the transformation
\begin{equation}
{d^2N \over dE_\nu d\Omega} \propto {d^2N \over dE_\nu d\cos\theta}
= {d^2N \over dE_\pi d\cos\theta^\star} 
J(E_\pi, \cos\theta^\star; E_\nu, \cos\theta)
\propto (E_p - E_\pi)^5 J,
\label{s12}
\end{equation} 
where the Jacobian is given by
\begin{equation}
J(E_\pi, \cos\theta^\star; E_\nu, \cos\theta)=
\left| \begin{array}{cc}
{\partial E_\pi \over \partial E_\nu} &
{\partial \cos\theta^\star \over \partial E_\nu} \\
{\partial E_\pi \over \partial \cos\theta} &
{\partial \cos\theta^\star \over \partial \cos\theta}
\end{array} \right|.
\label{s13}
\end{equation} 

The ``exact'' form of the Jacobian is somewhat lengthy, so we will simplify to
the extent we can by noting that when $E_\nu \gg m_\pi$, the parent pion has
$E_\pi \gg m_\pi$ also, and so $\beta_\pi \approx 1$.  Also, part 3 tells us
that $\theta$ is very small for any value of $\theta^\star$.

We already have relation (\ref{s14}) between $E_\nu$, $\tan\theta$
and $\sin\theta^\star$, so we can write
\begin{equation}
\cos\theta^\star = \sqrt{1 - \sin^2\theta^\star} \approx 
\sqrt{1 - {E_\nu^2 \over E_\nu^{\star 2}} \tan^2\theta}
= \sqrt{1 - {E_\nu^2 \over E_\nu^{\star 2}} \left( {1 \over \cos^2\theta} - 1
\right)} \, .
\label{s15}
\end{equation}
Thus,
\begin{equation}
{\partial \cos\theta^\star \over \partial E_\nu} \approx 
- { {E_\nu \over E_\nu^{\star 2}} \tan^2\theta \over
\sqrt{1 - {E_\nu^2 \over E_\nu^{\star 2}} \tan^2\theta}}
\approx 
- {E_\nu \theta^2 \over E_\nu^{\star 2} \cos\theta^\star}\, ,
\label{s16}
\end{equation}
for small $\theta$, and
\begin{equation}
{\partial \cos\theta^\star \over \partial \cos\theta} \approx 
{ {E_\nu^2 \over E_\nu^{\star 2} \cos^3\theta} \over
\sqrt{1 - {E_\nu^2 \over E_\nu^{\star 2}} \tan^2\theta}}
\approx 
{E_\nu^2 \over E_\nu^{\star 2} \cos\theta^\star}\, .
\label{s17}
\end{equation}

We can also use time components of eq.~(\ref{s8}) to write
\begin{equation}
\gamma_\pi = {E_\pi \over m_\pi} 
= {E_\nu \over E_\nu^\star (1 + \beta_\pi \cos\theta^\star)}
\approx {E_\nu \over E_\nu^\star (1 +  \cos\theta^\star)}
\label{s18}
\end{equation}
Hence,
\begin{equation}
{\partial E_\pi \over \partial E_\nu} \approx 
{m_\pi \over E_\nu^\star (1 +  \cos\theta^\star)}
- {m_\pi E_\nu \over E_\nu^\star (1 +  \cos\theta^\star)^2}
{\partial \cos\theta^\star \over \partial E_\nu}
\approx {E_\pi \over E_\nu} + {E_\pi^2 \theta^2 \over m_\pi E_\nu^\star
\cos\theta^\star}\, ,
\label{s19}
\end{equation}
and
\begin{equation}
{\partial E_\pi \over \partial \cos\theta} \approx 
- {m_\pi E_\nu \over E_\nu^\star (1 +  \cos\theta^\star)^2}
{\partial \cos\theta^\star \over \partial \cos\theta}
\approx 
-{E_\pi^2 E_\nu \over m_\pi E_\nu^\star \cos\theta^\star}\, .
\label{s20}
\end{equation}

The Jacobian (\ref{s13}) is therefore
\begin{equation}
J \approx 
\left| \begin{array}{cc}
{E_\pi \over E_\nu} + {E_\pi^2 \theta^2 \over m_\pi E_\nu^\star
\cos\theta^\star} &
- {E_\nu \theta^2 \over E_\nu^{\star 2} \cos\theta^\star} \\
-{E_\pi^2 E_\nu \over m_\pi E_\nu^\star \cos\theta^\star} &
{E_\nu^2 \over E_\nu^{\star 2} \cos\theta^\star}
\end{array} \right|
= {E_\pi E_\nu \over E_\nu^{\star 2} \cos\theta^\star}\, ,
\label{s23}
\end{equation} 
and hence the neutrino spectrum can be written from eq.~(\ref{s12}) as
\begin{equation}
{d^2N \over dE_\nu d\cos\theta} \propto (E_p - E_\pi)^5 
{E_\pi E_\nu \over \cos\theta^\star}\, .
\label{s24}
\end{equation} 

Because the factor $\cos\theta^\star$ in the denominator of the Jacobian can
go to zero, it is possible that the neutrino flux is higher for nonzero
values of the lab angle $\theta$.

\item
On the axis, $\theta = 0$, $\theta^\star = 0$, and 
$E_\pi = m_\pi E_\nu / 2 E_\nu^\star \approx 2 E_\nu$ according
to eq.~(\ref{s18}).  In this case, the neutrino spectrum (\ref{s24}) is
\begin{equation}
{d^2N(\theta = 0) \over dE_\nu d\cos\theta} \propto \left( E_p - 
{m_\pi E_\nu \over 2 E_\nu^\star} \right)^5 
E_\nu^2.
\label{s25}
\end{equation}
The peak of the spectrum occurs at
\begin{equation}
E_{\nu,\rm peak} = {4 E_\nu^\star \over 7 m_\pi} E_p \approx {E_p \over 8}\, .
\label{s26}
\end{equation}

\item
For an off-axis neutrino beam (at a nonzero value of angle $\theta$)
we must evaluate the spectrum (\ref{s24}) using relations (\ref{s15}) and
(\ref{s18}).  This is readily done numerically.  For example, a plot of
the pion energy $E_\pi$ needed to produce a neutrino of energy $E_\nu$ at
various angles $\theta$ is shown below.

\vspace{0.1in}
\centerline{\includegraphics*[width=4in]{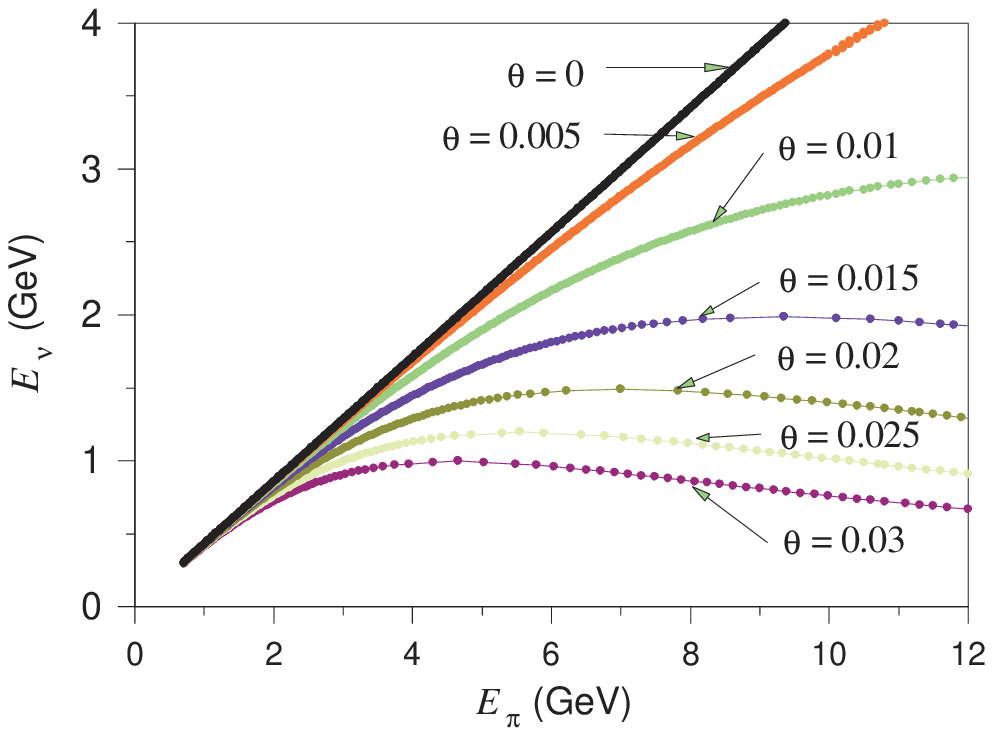}}

As expected from part 3,
we see that for a given angle $\theta$, there is a maximum possible neutrino
energy, and as the neutrino energy approaches this value, a large range of pion
energies contributes to a small range of neutrino energies.  This will result in an enhancement of the neutrino spectrum.  If we desire the
enhancement at a particular neutrino energy, we should look for the
neutrinos close to the angle $\theta_{\rm max}$ given in 
eq.~(\ref{s14a}), which is independent of the proton/pion energy.

A numerical evaluation of the neutrino spectrum (\ref{s24})
for several values of angle
$\theta$ with respect to the proton/pion beam is shown below.

\vspace{0.1in}
\centerline{\includegraphics*[width=4in]{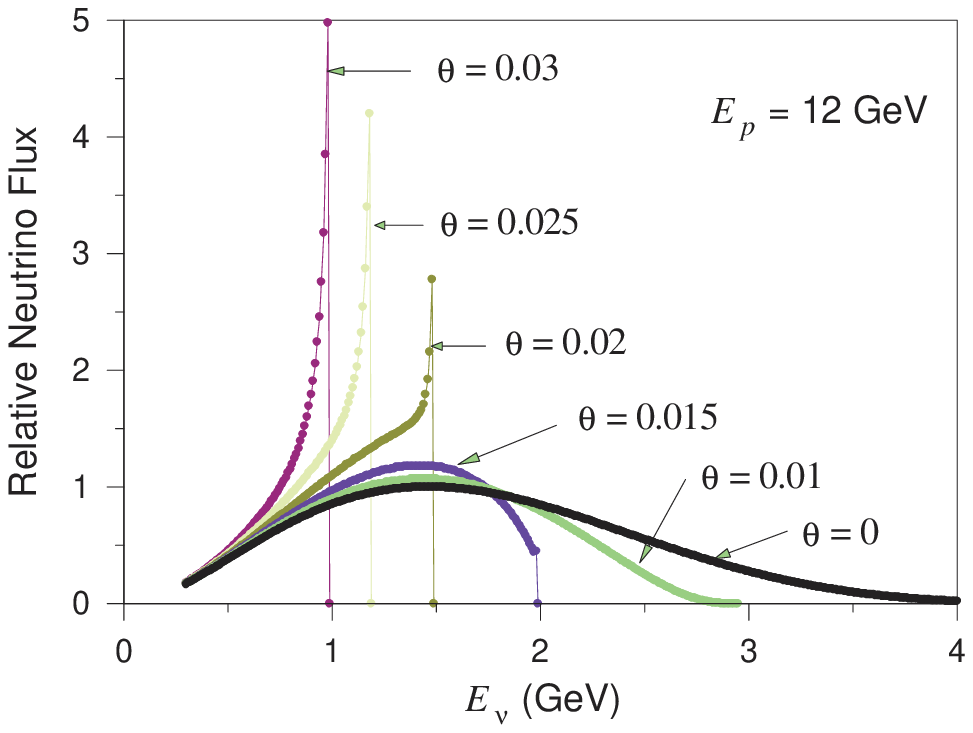}}

We see that the spectrum of neutrinos at a nonzero angle is peaked at
a lower energy, and is narrower, than that at zero degrees, 
due to the existence of
a maximum possible neutrino energy (\ref{s14c}) in decays at a given angle
to the direction of the parent pion.
This effect is especially prominent
when $E_{\nu,\rm max}(\theta) \approx (30$ MeV)/$\theta$ is less than
$E_{\nu,\rm peak}(\theta = 0)$, as then there is a substantial rate of
higher energy pions all of which decay into a narrow band of neutrino energies
at this angle.

The spectral narrowing of an off-axis neutrino beam remains
in more complete calculation \cite{e889,J2K}
that include the nonzero transverse
momenta of the pions before and after passing through the ``horn", although the spectrum will not
have such hard edges, and the favored angle-energy combination is
$\theta \approx (50$ MeV)/$E_\nu$.

In sum, the existence of a maximum energy for neutrinos that decay
at a given angle to their parent pions implies that many different
pion energies contribute to the this neutrino energy, which enhances the
neutrino spectrum at this angle-energy combination, 
$\theta \approx (30$-50 MeV)/$E_\nu$.

\end{enumerate}

\section{Acknowledgement}

The author thanks Fritz DeJongh, Milind Diwan, Debbie Harris, Steve Kahn,
Bob Palmer and Milind Purohit for useful discussions on neutrino beams, both
on- and off-axis.


\begin{thebibliography}{99}

\bibitem{Sternheimer}
R.M.~Sternheimer,
{\sl Energy Distribution of $\gamma$ Rays from $\pi^0$ Decay},
Phys.\ Rev.\ {\bf 99}, 277 (1955); for a pedagogic discussion of $\pi^0$
decay similar to that of the present note, see K.T.~McDonald,
{\sl Neutral Pion Decay} (Sept.\ 15, 1976), \hfill\break
http://puhep1.princeton.edu/\~mcdonald/examples/piondecay.pdf
 

\bibitem{e889}
D.~Beavis \etal,
{\sl Long Baseline Neutrino Oscillation Experiment, E889, Physics
Design Report},
BNL 52459 (April 1995), Chap.~3,\hfill\break
http://puhep1.princeton.edu/\~mcdonald/nufact/e889/chapter3a.pdf

\bibitem{J2K}
Y.~Itow \etal,
{\sl The JHF-Kamioka neutrino project},
hep-ex/0106019.



\end{thebibliography}
\end{document}